# Improved Long-Term Prediction of Chaos Using Reservoir Computing Based on Stochastic Spin-Orbit Torque Devices

Cen Wang[1], Xinyao Lei[1], Kaiming Cai[1,2], Xiaofei Yang[1], Yue Zhang[1,3*]

[1]School of Integrated Circuits, Huazhong University of Science and Technology, Wuhan, 430074, P. R. China

[2]School of Physics, Huazhong University of Science and Technology, Wuhan, 430074, P. R. China

[3]Hubei Key Laboratory of Micro-Nanoelectronic Materials and Devices, Hubei University, Wuhan 430062, China.

*Corresponding author: yue-zhang@hust.edu.cn (Yue Zhang)

**Abstract**

Predicting chaotic systems is crucial for understanding complex behaviors, yet challenging due to their sensitivity to initial conditions and inherent unpredictability. Probabilistic Reservoir Computing (RC) is well-suited for long-term chaotic predictions by handling complex dynamic systems. Spin-Orbit Torque (SOT) devices in spintronics, with their nonlinear and probabilistic operations, can enhance performance in these tasks. This study proposes an RC system utilizing SOT devices for predicting chaotic dynamics. By simulating the reservoir in an RC network with SOT devices that achieve nonlinear resistance changes with random distribution, we enhance the robustness for the predictive capability of the model. The RC network predicted the behaviors of the Mackey-Glass and Lorenz chaotic systems, demonstrating that stochastic SOT devices significantly improve long-term prediction accuracy.

Accurate chaos prediction is essential across various fields due to its ability to forecast future states in highly sensitive chaotic systems. This capability is crucial in scientific research (e.g., weather forecasting, ecological dynamics), engineering applications (e.g., machinery maintenance, control systems), financial markets (e.g., stock price prediction, risk management), and healthcare (e.g., disease spread, physiological signal analysis). However, chaos prediction still faces significant challenges such as sensitivity to initial conditions, data uncertainty, high computational complexity, multi-scale characteristics, and so on.[1-2]

Reservoir computing (RC)[3-4] is a framework for processing time series data using a recurrent neural network (RNN) with a fixed, randomly connected reservoir. The reservoir transforms input data into a high-dimensional space, capturing temporal patterns. Echo State Networks (ESNs) [Fig.1a],[5] a widely used form of RC, are well-suited for chaotic prediction due to their ability to handle nonlinear and complex dynamics, map input data to a high-dimensional space, and effectively process time series data.[6-7]

In general, noise is detrimental to accurate prediction in machine learning. However, in terms of the chaos prediction, noise performs beneficial advantages. Recent studies show that introducing noise into RC for chaotic prediction enhances model robustness and prevents overfitting, known as noise regularization. Noise can also increase the dynamic complexity of the reservoir, improving its ability to capture chaotic behaviors.[8-10]

Spintronic devices are ideal for RC hardware due to their nonlinear spin dynamics derived from the Landau-Lifshitz-Gilbert (LLG) equation.[11] In spintronics devices, Spin-Orbit Torque (SOT) devices are advantageous for their low power consumption (fJ/bit) and fast writing speed (ps/bit).[12,13] The Dzyaloshinskii-Moriya Interaction (DMI) at the heavy metal (HM)/ferromagnetic (FM) interface of an SOT device reduces the energy barrier and retention time of SOT-driven magnetization switching,[14,15] leading to continuous resistance changes with random distribution under thermal fluctuation [Fig. 1b-c].[16,17]

Very recently, it is demonstrated that SOT Magnetic-Tunnel Junctions (SOT-MTJs) are suitable for True-Random-Number Generators (TRNGs) and stochastic computing algorithms, such as Restricted Boltzmann Machines (RBM).[18,19] The switching probability of TRNGs based on SOT-MTJs can be precisely controlled, allowing for the generation of various distributions like Gaussian, uniform, and exponential.[20] These studies highlight the promising advantages of SOT-MTJ devices for implementing RBMs and TRNGs, particularly in terms of speed, power efficiency, and the ability to generate configurable random distributions. Given their nonlinear dynamics and stochasticity, SOT devices are potential candidates for chaos prediction by ESN [Fig.1d], however, the related researches have not been reported yet.

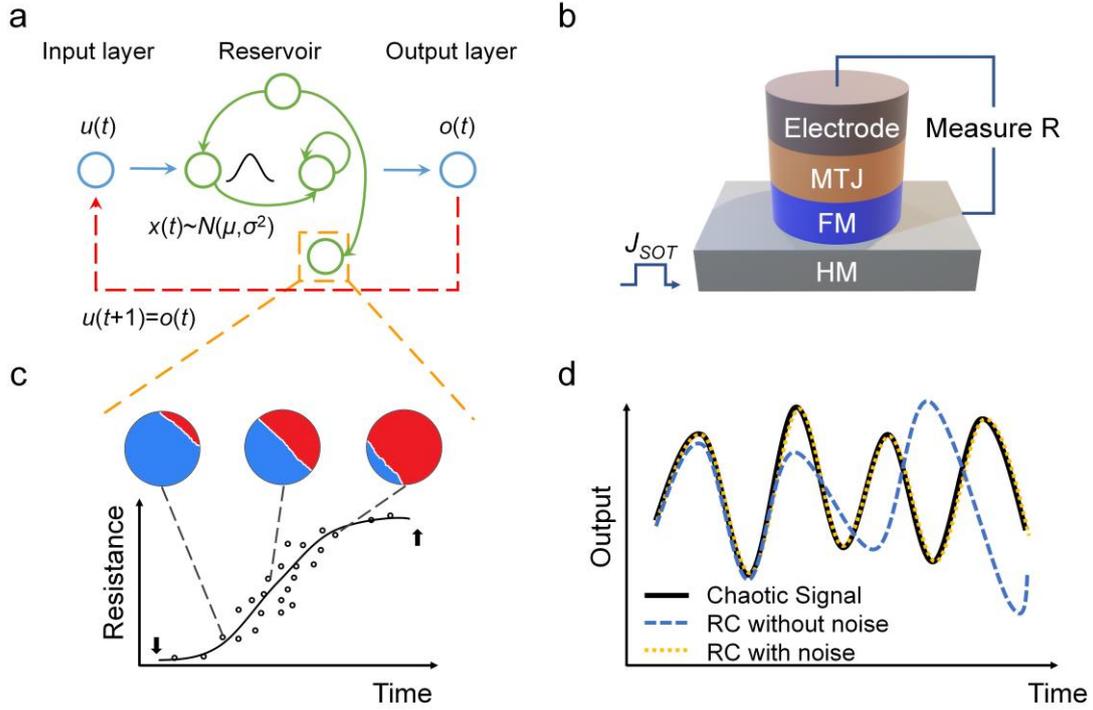

**Fig. 1 (a) Two configurations of ESN: open-loop configuration for training (original data like chaos dynamics as input signal) (blue lines) and autonomous evolution for prediction (the output as the input signal at next time step) (red dashed line). (b) SOT devices with an in-plane current, the resistance is measured by MTJ. (c) Variation of resistance distribution with current applying time. Bule and red areas imply the FM magnetization states along the direction of -z or +z, and the white line between them means the domain wall. (d) Example of the noise influence in the chaotic prediction by RC.**

In this work, we proposed an SOT device to achieve nonlinear resistance change with random distribution. We used it to simulate the reservoir in ESN to predict two well-known benchmark chaotic dynamics: Mackey-Glass time series[21] and Lorenz chaotic system.[22] The results showed that under the influence of noise in magnetic switching, the long-term chaotic prediction and the network robustness were highly improved as compared to the device without the consideration of noise.

We considered an ESN model as illustrated in Fig. 1a. The ESN consists of three components: the input layer, the reservoir, and the output layer. During training, the input layer receives external signals (blue lines in Fig.1a), which are fed into the reservoir—a large, sparsely connected, recurrent network with fixed, random weights that generate complex temporal dynamics. The output layer reads the reservoir states to produce the final output, and its weights are the only ones trained. After training, the loop of ESN will be closed (red line in Fig.1a), and the output of the ESN will be feedback to the input layer as the input signal at the next time step, so that the ESN generates complex dynamics autonomously.

The ESN updates according to the following equations:[23]

$$x(t+1) = (1-\alpha)x(t) + \alpha \tanh(W_{in} \cdot u(t+1) + W \cdot x(t)) \quad (1).$$
$$y(t+1) = W_{out} \cdot x(t+1)$$

In which, $\alpha$ is the learning rate, $u$ is the input signal, $x$ means the reservoir state, and $y$ is the output result. $W_{in}$, $W_{out}$, and $W$ represent the weight matrices connecting the input layer with reservoir, the reservoir with output layer, and the interconnecting within the reservoir, respectively.

We consider a nano-sized circular SOT device made of a FM/HM bilayer [Fig. 1(b)] to simulate the non-linear behavior of the ESN reservoir. The resistance variation of SOT behaves as a random distribution, and the mean value could be fitted as a tanh curve $f(\cdot)$, and Eq.(1) can be converted to $x(t+1) = (1-\alpha)x(t) + \alpha \cdot N(\mu,\sigma)$, where $\mu = f\left[W_{in} \cdot u(t+1) + W \cdot x(t)\right]$ (details of magnetic simulation and the curve fitting can be seen in method).

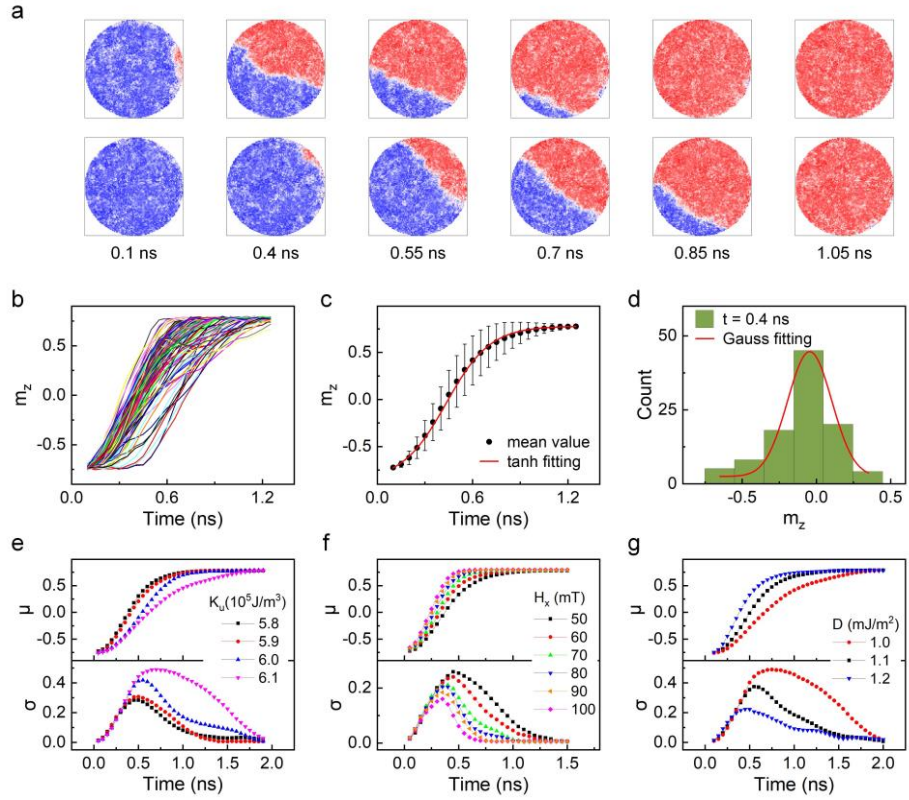

Fig. 2 (a) Snapshots of twice magnetization switching based on DW nucleation and motion under the same operations. The labels under the snapshots indicate the current applying time (ns). (b) Different $m_z$ variations with current applying time for 100 repetitions under the same simulation conditions. (c) Average $m_z$ variation (black dots) with standard deviation (black bars) and the tanh fitting curve (red line). (d) Frequency counts for $m_z$ at $t = 0.4$ ns conforms to a Gaussian distribution. (e-g) The variations of the mean value $\mu$ and standard deviation $\sigma$ of $m_z$ under different $K_u$ (e), $H_x$ (f), and $D$ (g). The parameters are set as $K_u=5.8\times10^5$ J/m$^3$ for (f) and $6.1\times10^5$ J/m$^3$ for (g), $H_x=50$ mT for (e) and (g), and $D=1.0$ mJ/m$^2$ for (e) and (f).

The stochastic magnetic switching relies on the DW generation and motion under current. As depicted in Fig.2a, under the injection of a current ($J = 4\times10^{10}$ A/m$^2$) along the $x$ direction, the DW forms at upper right corner of the disk and then moves diagonally, giving rise to gradual changes of the total magnetization. We extracted time-dependent average z-component magnetization ($m_z$), which is proportional to the resistance variation of the SOT device detected by an MTJ.

In an SOT device, thermal fluctuation and spatial disorder introduce stochasticity in the magnetic switching (Fig.2a), which can be seen from different $m_z \sim t$ curves for repeated operations under the same conditions by 100 times (Fig.2b). Due to varying velocities of the DW motion, the $m_z \sim t$ curve presents typical nonlinear behaviors. The mean of $m_z$ can be well fitted by a tanh curve (Fig.2c), and its frequency distribution conforms to a Gaussian distribution (Fig.2d).

To modify the stochasticity in temporal $m_z$, we adjusted the simulation parameters $K_u$, $H_x$ and $D$. Different nonlinear features and standard deviations are shown in Fig. 2 (e-g). We repeated the simulation 50 times under every set of parameters, and calculated the mean value $\mu$ and standard deviation $\sigma$ of $m_z$.

As depicted in Fig. 2e, as $K_u$ increases, $\mu$ varies slower at a larger $\sigma$. This is unexpected in a single-domain system depicted by a macro-spin model, in which a larger $K_u$ usually depresses the stochasticity from noise. In the multi-domain system, however, the increase of $K_u$ leads to a higher energy barrier for the DW motion, so that magnetic switching needs longer time and presents more susceptible to evolve the magnetic states with different configurations. Similarly, the increase of $H_x$ and $D$ also leads to the faster magnetic switching and the inhibition of $\sigma$ (Figs. 2f-g).

To demonstrate the ability of ESN with stochastic SOT devices to predict chaos, we selected two benchmark chaotic systems: Mackey-Glass chaotic system and Lorenz chaotic system. They both exhibit chaotic behavior, but differ significantly in their structures, predictability, and applications.

The Mackey-Glass (M-G) chaotic system is a one-dimensional system with a delay term. It was originally introduced by Michael Mackey and Leon Glass in 1977 to describe physiological control systems, such as blood production regulation.[21] The standard form of the M-G equation is:

$$\dot{X}(t) = \frac{\beta X(t-\tau)}{1+[X(t-\tau)]^n} - \gamma X(t) \qquad (2).$$

The parameters are typically set as $\beta=0.2$, $\gamma=0.1$, $n=10$, and $\tau=17$ to make the system chaotic. In this study, based on the original dataset from time-series data M-G by Mantas Lukosevicius,[24] we used 4500 samples to learn, in which 3000 for training and 1500 for prediction.

The Lorenz system, a three-dimensional chaotic system, was originally derived from simplified equations of atmospheric convection. The Lorenz system is defined by a set of three coupled, nonlinear differential equations:

$$\frac{dx}{dt} = \sigma(y-x)$$
$$\frac{dy}{dt} = x(\rho-z)-y \quad (3).$$
$$\frac{dz}{dt} = xy-\beta z$$

Here $\sigma$, $\rho$, and $\beta$ are the Prandtl number, Rayleigh number, and geometric factor. The system exhibits chaotic behavior for $\sigma = 10$, $b = 8/3$, and $r = 28$,[1] that we used in this study. We generated the samples by solving the dynamic equations using the fourth-order Runge Kutta algorithm with step 0.01. We only considered the $x$ coordinate of the chaotic system, which was rescaled by a factor of 0.02. The first 3000 transient values were discarded and the remaining 3000 for training and 500 for autonomous evolution.

We utilized different $\mu$ of SOT devices, with or without $\sigma$, to predict the chaotic evolution of the M-G system. In Lorenz system prediction, we chose one material parameter (fixed $\mu$ and $\sigma$) of the SOT device but changed a series of hyperparameters to train the ESN.

In both chaotic prediction tasks, $W$ and $W_{in}$ are randomly initialized. During the training, the sample data was fed to the input layer and the reservoir evolved as Eq. (1). $W_{out}$ was trained via ridge regression. For prediction, the loop was closed and the output was feedback to the input at the next step, so that the ESN evolves autonomously.

To quantitatively evaluate the predictions, we calculated the root mean square error (RMSE) to estimate the accuracy of the autonomous operation with the originate chaotic samples as $RMSE = \sqrt{\|y_o - y_p\|^2 / n}$, where $y_o$ and $y_p$ mean original sample data and predicted outputs, respectively, and $n$ is the length of the testing data. The smaller

*RMSE* means more accurate prediction. On the other hand, we utilized power spectrum density (PSD) to quantify the similarity between predicted dynamic behaviors and originate chaotic characteristic, and calculated the cross-correlation coefficient $\rho$ of the PSD of the evolutions and the original chaotic data.[25] The higher overlap of PSD and the larger $\rho$ imply that the autonomous evolution of ESN is more approximate to the chaotic dynamics.

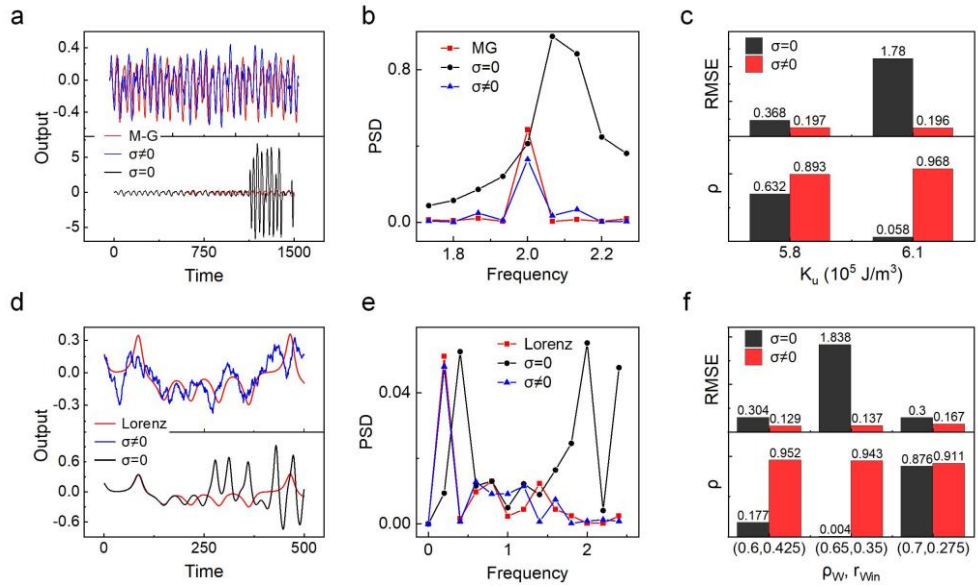

**Fig .3 (a) For M-G chaotic prediction, autonomous evolution of ESN based SOT devices with (upper) and without noise (below) for $K_u$=6.1×10$^5$ J/m$^3$. (b) PSD of the originate M-G data and ESN evolutions in (a). (c) RMSE of the ESN predictions with the originate M-G data (upper) and $\rho$ between the PSD of ESN evolutions with that of the M-G data (below) under different $K_u$. (d)~(f) The same operation for the Lorenz chaotic system. The hyperparameter for (d) and (e) is ($\rho_W$, $r_{Win}$)=(0.6, 0.425). Red, blue, and black lines in (a-b) and (d-e) imply the original chaotic data, prediction by ESN with noise, and prediction by ESN without noise, respectively.**

For the M-G system, we used SOT devices with $K_u$ = 5.8 and 6.1 × 10$^5$ J/m$^3$ to simulate the ESN reservoir (Fig.3a-c). The size of reservoir nodes $N$ is 1000, and the hyperparameters $\alpha$ (learning rate), $\rho_W$ (spectral radius of $W$), and $r_{Win}$ (random initialization range - $r_{Win} \sim + r_{Win}$ of $W_{in}$) were set as $(\alpha, \rho_W, r_{Win}) = (0.3, 0.8, 2.5)$ for $K_u = 5.8 \times 10^5$ J/m$^3$ and $(\alpha, \rho_W, r_{Win}) = (0.23, 0.85, 3.25)$ for $K_u = 6.1 \times 10^5$ J/m$^3$.

For the Lorenz prediction task, we used the SOT device with $K_u = 6.1 \times 10^5$ J/m$^3$ to simulate the ESN reservoir. We attempt three different initial hyperparameter conditions (all $N=1000$, $\alpha=0.3$) to train the ESN and predict the Lorenz chaotic dynamics as depicted in Fig .3(d-f).

For both tasks, the ESN was trained and predicted with (blue lines in Fig. 3) and without the randomness distribution (black lines in Fig. 3). The predicted results show that the reservoirs without noise are more prone to divergence ($RMSE > 1$), while those with noise provide better predictions, as indicated by the higher overlap of PSD (Figs. 3b and 3e), the smaller $RMSE$ values, and the larger $\rho$ (Figs. 3c and 3f). Meanwhile, for the long-term prediction, the introduction of noise significantly enhanced the robustness of the ESN, as the ESN without noise is more sensitive to the variation of both material parameters (Fig. 3c) and hyperparameters (Fig. 3f).

In this study, we have demonstrated the effectiveness of utilizing SOT devices for nonlinear resistance change with random distribution to improve RC systems, specifically for chaotic predictions. The incorporation of stochastic SOT devices into ESNs showcased remarkable improvements in the prediction accuracy and robustness of chaotic dynamics, as exemplified by the Mackey-Glass system and Lorenz chaotic system. Overall, our study provides a novel perspective on enhancing RC systems using nonlinear SOT devices with random distribution, offering a promising pathway for future research and practical implementations in chaotic prediction tasks.

**Methods**

**Micromagnetic Simulation:**

We simulated SOT driven magnetic switching dynamics by using "Object Oriented MicroMagnetic Framework (OOMMF)" software. It uses finite difference methods to solve the LLG equation as:[11]

$$\frac{d\vec{m}}{dt} = -\gamma\mu_0 \vec{m} \times \vec{H}_{eff} + \alpha(\vec{m} \times \frac{d\vec{m}}{dt}) + \vec{\tau}_{SOT} \quad (4).$$

Here $\vec{m} = \vec{M}/M_S$ is the normalized magnetization vector with $M_S$ the saturation magnetization; $\gamma$ is the gyromagnetic ratio of electrons; $\mu_0$ is the vacuum permeability; $\alpha$ is the damping constant; $\vec{H}_{eff}$ is the effective field including the contributions from exchange coupling, demagnetization energy, anisotropy energy, and DMI.

The thermal fluctuation is modeled as a stochastic effective field with white noise following the Gaussian distribution. The average stochastic field is zero, and the square of the standard deviation is:[16]

$$\sigma_{thermal}^2 = \frac{\alpha}{1+\alpha} \times \frac{2k_B T}{\gamma \mu_0 M_S V} \tag{5}$$

Here $V$ is the volume of the FM layer.

The damping-like SOT term is expressed by:

$$\vec{\tau}_{SOT} = \frac{\gamma \hbar \theta_{SH} J}{2|e|dM_s} \vec{m} \times \vec{\sigma} \times \vec{m} \tag{6}$$

Here $\hbar$ is the reduced Planck constant; $e$ is the electron charge; $d$ is the thickness of the FM layer; $\theta_{SH}$ is the spin Hall angle; $J$ is the current density in the HM layer. $\vec{\sigma}$ is the polarization direction of the spin. We took into account the disorder of magnetic parameters given by thickness fluctuations of the film. For simulations of disordered films, the sample is divided into grains of average area 400 nm² by Voronoi tessellation, each grain having a normally distributed random thickness $t_G = h + N(0,1)rh$, with $r$ the relative magnitude of the grain-to-grain thickness variations and $h$ the mean thickness of the sample. These thickness fluctuations are then modeled as $M_S^G = M_S t_G/h$ and $K_u^G = K_u h/t_G$.[26]

The nanodisk diameter is 200 nm, with a thickness of 0.8 nm. The simulated magnetic material is Pt/Co multilayers with perpendicular magnetic anisotropy, and the parameters are as follows: the cell dimension was 1 nm × 1 nm × 0.8 nm, $\theta_H = 0.1$, $M_S = 5.4 \times 10^5$ A/m,[27] $J = 4 \times 10^{10}$ A/m², $A = 1 \times 10^{-11}$ J/m, $\alpha = 0.27$,[28] $K_u = 5.8 \times 10^5$ J/m³, $r=3\%$, $H_x=50$ mT, and $D = 1.0$ mJ/m².

We repeated the simulations for 50 times under different $K_u$ and extracted $m_z$ variation with current applying time. The variation of the mean value $\mu$ of $m_z$ is fitted as:

$$\mu = a \cdot tanh(bx+c) - d \tag{7}$$

and $\sigma$ can be fitted as a function of $\mu$ by fourth order polynomial as:

$$\sigma = p_4\mu^4 + p_3\mu^3 + p_2\mu^2 + p_1\mu + p_0 \tag{8}$$

The fitting parameters are listed in TABLE 1.

**TABLE 1: Fitting parameters of $\mu$ and $\sigma$ of SOT devices under different $K_u$**

| $K_u$ (J/m³) | 5.8 × 10⁵ | 6.1 × 10⁵ | $K_u$ (J/m³) | 5.8 × 10⁵ | 6.1 × 10⁵ |
|---|---|---|---|---|---|
| a | 0.86 | 0.987 | $p_4$ | -0.5058 | -0.6843 |
| b | 3.29 | 1.607 | $p_3$ | -0.2428 | -0.2639 |
| c | 0.144 | 0.163 | $p_2$ | -0.1168 | -0.3476 |
| d | -0.09 | -0.20455 | $p_1$ | 0.1888 | 0.2242 |
|   |   |   | $p_0$ | 0.2532 | 0.4737 |

**Acknowledgements:**

The authors acknowledge the financial support from the National Natural Science Foundation of China (Grant No: U2141236) and the Department of Education of Hubei Province (Grant No: B2020333).